\documentstyle[revtex,eqsecnum]{aps}
\begin{document}
\draft
\begin{title}
The Last Three Minutes: Issues in Gravitational Wave
Measurements \\ of Coalescing Compact Binaries
\end{title}
\author{Curt Cutler,$^{(1)}$ Theocharis A.~Apostolatos,$^{(1)}$
Lars Bildsten,$^{(1)}$ Lee Samuel Finn,$^{(2)}$
Eanna E. Flanagan,$^{(1)}$ Daniel Kennefick,$^{(1)}$
Dragoljubov M. Markovic,$^{(1)}$ Amos Ori,$^{(1)}$ Eric Poisson,$^{(1)}$
Gerald Jay Sussman,$^{(1),*}$ and Kip S.~Thorne$^{(1)}$
}
\begin{instit}
$^{(1)}$Theoretical Astrophysics, California Institute of Technology, Pasadena,
CA 91125 \\
$^{(2)}$Department of Physics and Astronomy, Northwestern University, Evanston,
IL 60208
\end{instit}
\begin{abstract}
Gravitational-wave interferometers are expected to monitor the last
three minutes of inspiral and final
coalescence of neutron star and black hole binaries at distances
approaching cosmological, where the event rate may be many per
year.  Because the binary's accumulated
orbital phase can be measured to a fractional
accuracy $\ll 10^{-3}$ and relativistic
effects are large, the waveforms
will be far more complex,
carry more information, and be far harder to model
theoretically than has been
expected.  Theorists must begin now to lay a foundation for extracting
the waves' information.

\end{abstract}
\pacs{PACS numbers: 04.30.+x, 04.80.+z, 97.60.Jd, 97.60.Lf}
\narrowtext

A network of gravitational-wave interferometers (the American
LIGO \cite{abramovici}, the French/Italian VIRGO \cite{virgo} and
possibly others) is expected to be operating by the end of the
1990s.  The most promising waves
for this network are from the inspiral and
coalescence of neutron star and black hole
binaries \cite{300yrs,schutz_nature}, with an estimated event rate of
$\sim (3/$year)(distance/200 Mpc)$^3$ \cite{phinney_narayan}.

A binary's inspiral and coalescence will produce two gravitational
waveforms, one for each polarization.  By cross-correlating the
outputs of 3 or more interferometers, both
waveforms can be monitored and the source's direction can be
determined to within
$\sim 1$ degree \cite{schutz_nature,angular_resolution}.

We shall divide each waveform into two parts: the {\it inspiral
waveform}, emitted before tidal distortions become noticeable
(3 or fewer orbital cycles before complete
disruption or merger \cite{bildsten_cutler}),
and the {\it coalescence waveform}, emitted
during distortion, disruption, and/or merger.

As the binary, driven by gravitational radiation reaction, spirals
together, its {\it inspiral waveform} sweeps upward in
frequency $f$.  The interferometers will observe the last
several thousand cycles of inspiral (from $f \sim 10$ Hz to $\sim 1000$
Hz), followed by coalescence.

Theoretical calculations of the waveforms are generally made using the
Post-Newtonian (``PN'') approximation to general relativity.  Previous
calculations have focused on the Newtonian-order waveforms
\cite{schutz_nature,300yrs,wahlquist,abramovici} and
on PN modulations of their amplitude and frequency \cite{lincoln_will}.

Two of us \cite{cutler_finn} recently realized that
the PN modulations are far less important than
PN contributions to the secular growth of the waves'
phase $\Phi = 2\pi \int
fdt$ (which arise, largely, from PN corrections to
radiation reaction.)  The binary's parameters are determined
by integrating the observed (noisy) signal against theoretical
templates \cite{finn}, and
if the signal and template lose phase with each other by
one cycle (out of thousands) as the signal sweeps through the
interferometers' band, their
overlap integral will be strongly reduced.  Correspondingly, one can
infer each of the system's parameters $\lambda_i$ to an accuracy that is
roughly the change $\Delta \lambda_i$ which alters by unity the
number of cycles ${\cal N}_{\rm cyc}$ spent in the interferometers' band.

The only parameters $\lambda_i$ that can influence the inspiral template
significantly are
(i) the initial orbital elements, (ii) the bodies' masses
and spin angular
momenta, and (iii) if the spins are very large,
the bodies' spin-induced quadrupole moments.
(The quadrupole moments produce orbital phase shifts a little larger than
unity at $f \sim 10$ Hz, but negligible phase
shifts later \cite{bildsten_cutler}.  Below we shall
ignore this tiny effect.)

If (as almost always is the case) the binary's
orbit has been circularized by radiation reaction, then
the number of cycles spent in a logarithmic interval of
frequency, $d{\cal N}_{\rm cyc}/d \ln f = (1/2\pi )(d\Phi/d\ln f)$, is:
\FL
\begin{eqnarray}
{d{\cal N}_{\rm cyc}\over d\ln f} &=& {5\over96\pi} {1\over \mu
M^{2/3}(\pi f)^{5/3}} \left \{ 1 + \left ( {743\over336} +
{11\over4}{\mu\over M}\right ) x \right. \nonumber \\
&&- \left. [4\pi + {\rm S.O.}] x^{1.5} + [{\rm S.S.}]x^2 +
O(x^{2.5}) \right \}.
\label{N_cyc}
\end{eqnarray}
Here $M$ is the binary's total mass, $\mu$ its reduced mass, and
$x \equiv (\pi Mf)^{2/3} \simeq M/D$ the PN expansion
parameter (with $D$ the bodies' separation and $c=G=1$).  The  PN
correction [$O(x)$ term] is from \cite{wagoner_will}.  In the
P$^{1.5}$N correction, the
$4\pi$ is created by the waves'
interaction with the
binary's monopolar gravitational field as they propagate
from the near zone to the radiation zone \cite{poisson}, and the
``S.O.'' denotes contributions
due to spin-orbit coupling \cite{sussman_thorne_apostolatos}.  In the
P$^2$N correction the ``S.S.'' includes
spin-spin coupling effects
\cite{sussman_thorne_apostolatos} plus an
expression quadratic in $\mu/M$.

Since the leading-order, Newtonian contribution to Eq.~(\ref{N_cyc})
gives several thousand cycles of oscillation in the interferometers'
band, to be measureable the higher-order corrections
need only be as large as a part in several thousand, even when the
amplitude signal to noise ratio has only its typical value of
8.  It is this that
gives the waves' phasing its high potential accuracy.

The determination of the binary's parameters (masses,
spins, orbital elements) is made possible
by the various frequency dependences in Eq.~(\ref{N_cyc}).
Analytic \cite{cutler_flanagan,chernoff_finn} and Monte Carlo \cite{cutler}
calculations show that (i) the
``chirp mass'' $M_c \equiv \mu^{3/5}M^{2/5}$ [which governs the Newtonian
part of (\ref{N_cyc})] will typically be measured
to $0.1$ per cent, and (ii) {\it if} we
somehow knew that the spins were small, then the
reduced mass $\mu$ would be measured to
$\sim 1$ per cent for NS/NS and NS/BH binaries, and
$\sim 3$ per cent for BH/BH binaries.  (Here and below NS means a
$\sim 1.4 M_\odot$
neutron star and BH means a $\sim 10 M_\odot$
black hole.)  Unfortunately, the frequency ($x$) dependences
of the various terms in Eq.~(\ref{N_cyc}) are not sufficiently different
to give a clean separation between $\mu$ and the spins.
Preliminary estimates \cite{cutler}, in which the S.O. term in
Eq.~(\ref{N_cyc}) was taken into account but not the S.S. term,
suggest that the spin$/\mu$ correlation will
worsen the typical accuracy of $\mu$ by a factor
$\sim 20$, to $\sim 20$ per cent for NS/NS, $\sim 25$ per cent for NS/BH, and
$\sim 50$ per cent for BH/BH.  These worsened accuracies might be improved
significantly, however,
by the waveform modulations \cite{lincoln_will}---most especially,
modulations caused by
spin-induced precession of the orbit
\cite{sussman_thorne_apostolatos}.  Much
additional theoretical work is needed
to pin down the measurement accuracies for $\mu$ and
the spins.

The highest accuracy parameter information will come
from low frequencies where most of the waves' cycles are located
\cite{cutler_flanagan,chernoff_finn,cutler}.
(It thus is important that
the interferometers achieve good low-frequency performance).  When
the waves have swept to higher frequencies, the binary's parameters
may be fairly well known, and the subsequent, highly relativistic waveforms
might be used to map out the binary's innermost
spacetime geometry near the orbital plane.  It would
be useful to develop a parametrized, ``PPN-like'' multi-theory formalism
for translating the measured waveforms
into spacetime maps and for comparing those maps with the predictions of
various gravitation theories.

To make optimal use of the interferometers' data will require
general-relativity-based
waveform templates whose phasing is correct to within a half cycle or so
during the entire frequency sweep from $\sim 10$ Hz to $\sim 1000$ Hz.
By examining an idealized limit ($\mu \ll M$ and vanishing spins),
several of us \cite{poisson,cutler_finn_poisson_sussman} have discovered
that, {\it to compute
the templates with the desired accuracy via PN methods will be very
difficult}.  We calculated the waves from such a binary
to high accuracy using
the Teukolsky \cite{teukolsky} black-hole perturbation formalism
and then fit a PN expansion to the results, to obtain
\cite{cutler_finn_poisson_sussman}
\FL
\begin{eqnarray}
&&{d{\cal N}_{\rm cyc}\over d\ln f} = {f^2dE/df \over dE/dt} =
{5\over96\pi} {M/\mu\over x^{2.5}} \times
\nonumber \\
&&\times \frac{1 - {3\over2}x - {81\over8}x^2 -{675\over16}x^3+ ...
}{1 - {1247\over336}x + 4\pi x^{1.5} - 4.9x^2 - 38 x^{2.5} + 170 x^3 +
...} .
\label{N_cyc_RW}
\end{eqnarray}
Here
$dE/df$ is half the ratio of the changes of orbital energy $dE$ and orbital
frequency $d(f/2)$ as
the orbit shrinks, and $dE/dt$ is the
power carried off by gravitational waves.
The coefficients in the denominator after the $4\pi$ were obtained from
the fit; all other coefficients are known analytically.  The
coefficient 4.9 has accuracy $\sim 2$\% ; the 38,
$\sim 10$\%
; and the 170, $\sim 25$\%.

We can use Eq.~(\ref{N_cyc_RW}) to estimate how the phase accuracy
depends on the order
to which the PN expansion is carried.  If the template were computed
through P$^2$N order [all $x^{2.5}$ and $x^3$ terms in
Eq.~(\ref{N_cyc_RW}) omitted], then the phase error typically would reach a
half cycle when the frequency had swept from 10 Hz to only $\sim 15$ Hz,
and the total phase error from $10$ Hz to $1000$ Hz would
typically be $\sim 6$ cycles.  For
interferometers whose peak sensitivity is at $70$ Hz \cite{abramovici}, one
could use such
templates for a maximum range of $\sim 50$ Hz to $\sim 90$ Hz without
accumulating phase errors larger than a half cycle.  If the template were
computed through P$^{2.5}$N order, there would be no substantial
improvement.  It is not at all clear how far beyond P$^{2.5}$N the
template must be carried to keep the total phase error
below a half cycle over the entire range from $\sim 10$ Hz to
$\sim 1000$ Hz.

This slow convergence of the PN expansion might be improved
by cleverness in the way one expresses the expansion
(e.g., by the use of Pad\'e approximates).  However,
even with great cleverness and fortitude,
PN templates might never cover the entire inspiral range,
from $\sim 10$ Hz to $\sim 1000$ Hz. Evidently,
new techniques are needed for computing templates to
higher accuracy.  Two such techniques look promising:  A
``post-Teukolsky expansion,'' and a ``weak-reaction expansion.''

{\it The post-Teukolsky expansion} would expand in powers of
$\mu/M$ and thus would be useable only for a light body
orbiting a much heavier black hole.  The expansion's first step
would be the unperturbed hole's Kerr metric; the
second, the
Teukolsky formalism \cite{teukolsky} for the light body's first-order
perturbations;
and the third, incorporation of radiation
reaction \cite{galtsov}.  Each of these first three steps is now in hand,
though studies
of the consequences of the radiation reaction are only
beginning \cite{ori_etal}.  That the expansion
must be carried well beyond these three steps is evident
from the magnitude of the $\mu/M$ term in Eq.~(\ref{N_cyc}),
and from recent studies \cite{kidder} of the
influence of a finite $\mu/M$ on the last stable circular orbit.

{\it The weak-reaction expansion} would be a variant of numerical
relativity in which one expands in powers of $1/Q
\equiv (dE/dt) (\pi fE)^{-1}$ (a
measure of the effects of radiation reaction during one
orbit). Because $1/Q \sim x^{2.5}$ (with $x$ the PN
parameter)
is always $\ll 1$,
this expansion might produce adequate inspiral templates for
all binaries.  The expansion's first
step might be a numerical solution of
the Einstein equations
for the binary's metric, with standing-wave
boundary conditions at infinity and at the BH horizon(s) (if
any) \cite{detweiler}.  In
co-rotating coordinates, the equations would be
elliptic and
might be solvable to high accuracy by relaxation
techniques.  The second
step might be to switch from standing-wave to outgoing- and
downgoing-wave boundary conditions, and evaluate the resulting linearized
metric perturbations, and with them
the leading contributions to the
waveforms.  The third step might be the leading
effects of radiation reaction, including
the orbital inspiral.  To obtain templates with the desired
phase accuracy, one would have to carry the
expansion beyond this third step.

Although an optimally-precise {\it measurement} of the binary's parameters
requires templates faithful to general relativity's
predictions, a near optimal {\it search} for
binary-inspiral waves can use
a family of templates that merely span the range of expected waveform
behaviors, without being closely
related to the true general relativistic waveforms
\cite{cutler_flanagan_sussman}.  Theorists
need to develop a set of such {\it search templates} and optimize their
efficiency for wave searches.

Turn, now, from a binary's inspiral waveforms to its {\it coalescence
waveforms}.
By the beginning of coalescence, the binary's
parameters (masses, spins, geometry) will be known with fair accuracy,
and from its masses the nature of its bodies (BH vs NS) should be fairly
clear.  The coalescence waveforms can then be used to probe
issues in the physics of gravity and atomic nuclei:

If the bodies are BH's of comparable mass (and
especially if their spins are large and
not perpendicular to the orbital plane), then the coalescence
should produce
large-amplitude, highly nonlinear vibrations of spacetime
curvature that may reveal aspects of gravity
we have never seen.  Although supercomputer
simulations of such coalescences
are being attempted, they are so difficult,
especially for spinning holes, that
they may well {\it not} give definitive results without
observational guidance.  Conversely, the
observed waveforms will be very hard to interpret without the
guidance of simulations.

The interferometer measurements discussed
above may well be achieved in the early
years of LIGO/VIRGO.  Most of the measurements
discussed below are more difficult, and may require
more mature interferometers.

For a NS/BH binary with the BH spinning moderately fast, the NS
should disrupt tidally before plunging into the BH.
The NS disruption should be quick, as should the final
coalescence of a NS/NS binary: within about
one orbit, the NS/BH and NS/NS binaries may
get smeared into roughly axially symmetric configurations,
thereby shutting off their waves
\cite{rasio_shapiro,bildsten_cutler}.  The rapid wave shutoff and
contamination by laser shot noise
may prevent the coalescence waveforms from revealing more than
two numbers:  the maximum frequency $f_{\rm
max}$ reached, and the total wave energy $E_{\rm
GW}$ in the last $\sim 0.02$ seconds
(when the energy
emission is maximal).  Either number, however, would be
valuable.  It could tell us the NS radius $R_{\rm NS}$
(or a combination of the two NS radii)
\cite{flanagan_markovic_thorne}.  Since the NS
masses will already be known from the inspiral
waveforms, such measurements on a number of NS's could reveal the NS
radius-mass relation $R_{\rm NS}(M_{\rm NS})$, from which one can infer
the equation of state of
matter at densities from $\sim 1$ to 10 times nuclear \cite{lindblom}.

In preparation for the interferometers' measurements of
$f_{\rm max}$ and/or $E_{\rm
GW}$, theorists need to model tidal disruption in NS/BH binaries
and coalescence of NS/NS binaries, to determine
the dependence of $f_{\rm max}$ and
$E_{\rm GW}$ on $R_{\rm NS}$ and the binary's other
parameters.
(Such modeling is also driven by the possibility that these events are
the sources of observed gamma-ray bursts \cite{gamma}.)
Simple arguments \cite{bildsten_cutler,flanagan_markovic_thorne}
suggest that $f_{\rm max}$ will be
roughly $(1/2\pi)(M_{\rm NS}/R_{\rm NS}^3)^{1/2}$, so for a $1.4M_\odot$ NS,
a 15 km radius will lead to $f_{\rm max} \sim 1000$ Hz,
while a 10 km radius will lead to $f_{\rm max} \sim 2000$ Hz.
Similarly, $E_{\rm GW}$ will be proportional to $1/R_{\rm NS}$ so a
10 km NS will
emit $\sim 50$ per cent more energy just before disruption than will a
15 km NS.

To measure the shutoff frequency
$f_{\rm max}\sim 1000$ Hz in the midst of the interferometers' shot
noise will probably require specially configured (``dual recycled''
\cite{meers}) interferometers
that operate over adjustable, narrow bands $\Delta f$ around
adjustable central frequencies $f_o$.  Such interferometers, operated in
concert with broad-band ones, could give simple
``yes/no'' answers as to whether the waves swept through their
frequency bands.  By collecting such data
on a number of binaries, with various choices for the
frequencies $f_o$, one might zero in on $f_{\rm max}$ for
various types of binaries, and thence on the NS
radius-mass relation. The dual-recycled
sensitivities required for such a program, however, will be
difficult to achieve \cite{flanagan_markovic_thorne}.

Even more difficult, but not totally hopeless, will be the measurement
of $E_{\rm GW}$ via the waves' ``Christodoulou
memory'' \cite {christodoulou,flanagan_markovic_thorne} (a
slowly changing mean of the waveforms, produced
by the $1/r$ gravitational field of the emitted gravitons).  When the
waves' mean is monitored by an optimally filtered, broad-band
interferometer, it gets averaged over a time $\sim 0.02$
seconds, so the maximum level it reaches is proportional to $E_{\rm
GW}$.  Unfortunately, to measure the memory's level and thence
$E_{\rm GW}$ with even 50 per cent accuracy for binaries at 200 Mpc
distance is likely to require interferometers modestly better than
LIGO's so-called ``advanced detectors'' \cite{abramovici}.

Turn, next, from a binary's coalescence to the use of its waveforms as
cosmological probes.  Schutz
\cite{schutz_nature} has pointed out that a binary's distance from Earth
(more precisely, its ``luminosity distance'' $r_L$)
can be inferred from its inspiral waveforms, and that such
gravitationally measured distances,
when combined with electromagnetically measured redshifts $z$, might
determine the Hubble constant $H_o$ to better
precision than heretofore.  However,
this will require identifying electromagnetically the clusters of
galaxies in which the binaries lie, at least in some statistical
sense---a difficult task \cite{schutz_nature}.

Remarkably, for NS/BH binaries at
$z\agt1$, it may be possible to determine $z$ as well as $r_L$ from the
gravitational waveforms, and then
from $r_L(z)$ one might
infer the Universe's Hubble constant
$H_o$, mean density $\bar\rho$, and cosmological constant
$\Lambda$ \cite{markovic}.  (A variant of the method could also work for
NS/NS binaries, but NS/BH binaries emit more strongly and thus
can be seen to greater distances, making them
more promising than NS/NS.)

The key to measuring $z$ is the expectation (based on radio
pulsar observations) that NS masses will cluster around
$1.4M_\odot$.  (If this fails, but there is a
sharp cutoff in NS masses at some upper limit, then that limiting
mass can be used as the key to the measurements.)
For a binary at $z \agt 1$, the waveforms will be measured
as functions of {\it redshifted} time, and correspondingly
they will reveal redshifted
masses $(1+z)M_{\rm NS}$ and $(1+z)M_{\rm BH}$.  From the measured
luminosity distance and
redshifted masses, it should be fairly clear whether the binary is
NS/BH or not; and if so, then one can infer a ``candidate
redshift'' for the binary, $1+z_{\rm cand} \equiv (1+z)M_{\rm
NS}/1.4M_\odot$.  If the NS masses cluster
around $1.4M_\odot$, then a large sample of measured
distances and candidate redshifts should cluster around the true
distance-redshift relation.

This gravitational method of determining cosmological parameters might
be much more immune to evolutionary and ill-understood systematic
effects than are electromagnetic methods, because gravitational waves are
immune to absorption and scattering, and neutron star masses
might not be sensitive to the Universe's evolution.

A detailed analysis by one of us \cite{markovic} suggests that this
gravitational approach to cosmology might begin to bring useful information
when broad-band interferometers reach the LIGO ``advanced detector''
sensitivities \cite{abramovici}: with one year of observational data,
the one-sigma accuracies might be $\delta H_o \sim 0.01 H_o$, $\delta
\bar \rho \sim 0.1 \rho_{\rm crit}$, and $\delta\Lambda \sim 0.3 H_o^2$,
where $\rho_{\rm crit}$ is the critical density to close the Universe.

We conclude by noting that some of the issues discussed in this Letter have
implications for a possible future space-based
interferometer called LAGOS, which would operate in the band
0.0001--0.03 Hz \cite{bender}.  We shall discuss those implications
elsewhere.

For helpful discussions we thank David Chernoff, Larry Kidder, Andrzej Krolak,
E. Sterl Phinney, Bernard F. Schutz, Clifford M. Will, and
Alan Wiseman.  This research was supported in part by National
Science Foundation Grant PHY-9213508, and, in view of its
applications to LAGOS, by NASA grants NAGW-2897, 2920, and 2936, and by
a Weingart Foundation, Lee A. Dubridge Fellowship to L. Bildsten and an
Alfred P. Sloan Foundation Fellowship to L.~S. Finn.


\begin{references}
\bibitem{ori_etal}Two of us
(A. Ori and D. Kennefick) have recently proved
that ``quasi-circular'' orbits (orbits of constant Boyer-Lindquist
radial coordinate $r$) remain quasicircular as they shrink under
the influence of radiation reaction.
\end{references}
\end{document}